\RequirePackage{lineno}
\documentclass[aps,prl,twocolumn,superscriptaddress,groupedaddress,preprintnumbers]{revtex4-2}  % for review and submission
\usepackage{graphicx}  % needed for figures
\usepackage{dcolumn}   % needed for some tables
\usepackage{bm}        % for math
\usepackage{amssymb}   % for math
\usepackage{multirow}
\usepackage{epsfig}
\usepackage{amsmath}
\usepackage[colorlinks, linkcolor=blue]{hyperref}
\usepackage{ulem}
\usepackage{float}

\begin{document}

%\modulolinenumbers[3]
%\linenumbers
\lefthyphenmin=2
\righthyphenmin=2

\widetext
\hfill {MSUHEP-24-019}

\title{Improved Hessian Method in Global Analysis of Parton Distribution Functions}
\affiliation{Department of Modern Physics, University of Science and Technology of China, Jinzhai Road 96, Hefei, Anhui 230026, China}
\affiliation{Department of Physics and Astronomy, Michigan State University, East Lansing, MI 48823, USA}

\author{WenXiao Zhan} \affiliation{Department of Modern Physics, University of Science and Technology of China, Jinzhai Road 96, Hefei, Anhui 230026, China}
\author{Siqi Yang}\email{yangsq@ustc.edu.cn}
\affiliation{Department of Modern Physics, University of Science and Technology of China, Jinzhai Road 96, Hefei, Anhui 230026, China}
\author{Minghui Liu} \affiliation{Department of Modern Physics, University of Science and Technology of China, Jinzhai Road 96, Hefei, Anhui 230026, China}
\author{Liang Han} \affiliation{Department of Modern Physics, University of Science and Technology of China, Jinzhai Road 96, Hefei, Anhui 230026, China}
\author{Daniel Stump}\affiliation{Department of Physics and Astronomy, Michigan State University, East Lansing, MI 48823, USA}
\author{C.-P. Yuan} \affiliation{Department of Physics and Astronomy, Michigan State University, East Lansing, MI 48823, USA}

%\date{Oct. 11, 2017}
%\maketitle
\begin{abstract}
The Hessian method is widely applied in the global analysis of parton distribution functions (PDFs), 
which uses a set of orthogonal eigenvectors to give predictions of a physical observable. Its uncertainty 
is estimated based on the assumption that all physical observables 
can be approximately expressed as linear functions of the non-perturbative parameters in PDF. 
In this article, we report an improved Hessian method which takes the non-linear effects into account 
in the uncertainty estimation. 
A pseudo global analysis is designed to numerically test the new method. The non-linear uncertainties 
can significantly enlarge the original linear ones. Such uncertainties can be 
reduced with high precision data introduced into the global analysis. However, the non-linear effect could 
still be sizable corresponding to the current precision of the experimental measurements.
\end{abstract}
%\pacs{12.15.-y, 12.38.-t, 13.85.-t, 21.60.-n}
\maketitle

\section{I. Introduction}

In the past decades, a global analysis method based on quantum chromodynamics (QCD)
has been established to study the proton structure at high energy scale. It sufficiently analyzes
comprehensive data results from the deep inelastic scattering (DIS) experiments with the energy scale
of a few GeV, to the processes at hadron colliders at $\mathcal{O}(100)$ GeV.
The global analysis is expected to not only give the central predictions on
parton densities corresponding to the best fit of the data, but also provide a way that
the uncertainties can be easily calculated and extrapolated for any particular physical observable.
The Hessian method~\cite{Hessian} is one of the most widely used strategies to
estimate the PDF uncertainties, such as in the latest CT18NNLO and MSHT20 PDFs~\cite{CT18, MSHT20}. 
In this approach, a group of error PDF sets
are provided, which represent the possible variations on the
parameters of the non-perturbative functions of the quarks and gluons. These error sets correspond
to orthogonal eigenvectors in the original parameter space, so that the uncertainty of a particular physical
observable can be calculated using the error sets without complicated analysis on the correlations between
the original parameters.

In the current Hessian method, the uncertainty estimation is established as a linear approximation. 
Firstly, the orthogonal eigenvectors and their corresponding uncertainties are determined 
by assuming the physical observables of those data used in the global analysis 
can be expressed as linear functions of the non-perturbative PDF parameters. 
Secondly, when computing the PDF-induced uncertainty on a given physical observable 
using the eigenvectors, the observable is also assumed as linear function of 
the eigenvectors.  
The non-linear effects, appearing as the higher order derivatives in the expansion 
of physical observables as functions of PDF parameters or eigenvectors, are ignored. 
Such effect could be significant in high energy interactions, such as 
the LHC which provides precise experimental results in today's PDF 
global analysis.
A few previous studies have been performed on the non-linearities in the 
global analysis. For example, Ref.~\cite{epump} discussed contributions from the 
second-order terms based on the CT14 PDFs~\cite{CT14}, 
but only for those appearing as the diagonal 
elements in the Hessian matrix (as to be introduced later). 
Based on the same PDFs, it was found later in Ref.~\cite{CMSdijetRwt} that the diagonal 
higher order terms could be sizable in the analysis of the CMS di-jet measurement. 

In this work, we introduce an improved Hessian method, in which 
the non-linear effects are estimated as additional uncertainties. 
To make a detailed comparison between the improved method and the original one, a pseudo global fit is 
studied. As will be demonstrated in this test, the 
uncertainty on parton densities could be noticeably different from the original linear estimation. Such 
difference can be reduced as the precision of the input data in the global analysis improves. However, 
the needed precision to reduce the non-linear uncertainties to negligible level is far beyond 
the current precision of experimental observations. 
%In this work, we focus on the general method discussion. 
%For a real global analysis, the detailed strategy might be adjusted according 
%to the behavior of the particular data used in the fit.  

The article will be arranged as following: In section II, we review the general strategy of Hessian method; 
In section III, we discuss the non-linear effects in the global analysis; In section IV, 
we introduce the improved method 
to estimate the non-linear uncertainties; In section V, we perform a pseudo global analysis study to 
numerically test the improved method, and summarize the conclusion of this article.

\section{II. The original Hessian method}

In this section, we review the original Hessian method~\cite{Hessian}
and its linear uncertainty estimation. 
In Hessian method, a series of non-perturbative functions are introduced to describe 
the parton densities at the scale of $Q_0$. Taking from the CT18NNLO PDF as an example, 
the functions are written in the formalism as:

\begin{footnotesize}
\begin{eqnarray}\label{eq:npfunction}
  f_q(x, Q_0) = a_0 x^{a_1-1} (1-x)^{a_2} P_q(x; a_3, a_4, \cdot\cdot\cdot)
\end{eqnarray}
\end{footnotesize}

\noindent where $x$ is the momentum fraction of the parton $q$ inside the proton. $P_q$ is a polynomial designed to allow the parton density
to change its shape in order to
match the data inputs in different $x$ regions. $\{a_i\}$ are the original parameters being determined in the
global fit, by minimizing the $\chi^2$ defined as:

\begin{footnotesize}
\begin{eqnarray}
 \chi^2 = \sum_\alpha \frac{[D_\alpha - T_\alpha(\{a_i\})]^2}{\sigma^2_\alpha}
\end{eqnarray}
\end{footnotesize}

\noindent where $D_\alpha$ and $\sigma_\alpha$ are the experimental inputs and their corresponding uncertainties.
$T_\alpha(a_i)$
is the theory prediction of $D_\alpha$, depending on the PDF shape parameters $\{a_i\}$.
Given the fact that the best fit values of $\{a_i\}$ appear near the minimal $\chi^2$, we have:

\begin{footnotesize}
\begin{eqnarray}\label{eq:originalchi2}
  \chi^2 = \chi^2_\text{min} + \frac{1}{2}\sum_{i,j} \frac{\partial^2 \chi^2}{\partial a_i \partial a_j} (a_i - a^0_i) (a_j - a^0_j) + \cdot\cdot\cdot.
\end{eqnarray}
\end{footnotesize}

\noindent The Hessian matrix is then defined as $H_{ij} = \partial^2 \chi^2 / \partial a_i \partial a_j$.
Note that $\{ a_i\}$ are quite independent as free parameters 
in the non-perturbative functions, but their uncertainties 
in a global analysis can be correlated because experimental observables usually relate to multiple 
PDF shape parameters. The correlation depends on the particular set of experimental inputs used 
in a global fit.
One can convert $\{a_i\}$ to another set of orthogonal parameters $\{z_i\}$ by
diagonalizing $H_{ij}$:

\begin{footnotesize}
\begin{eqnarray}\label{eq:Hmatrix}
  \sum_j H_{ij} v_{jk} &=& \epsilon_k v_{ik}, \nonumber
\end{eqnarray}
\end{footnotesize}

\noindent where $\epsilon_k$ are the eigenvalues and $v_{ik}$ are the orthogonal and complete eigenvectors. Then, the displacements of $\{a_i\}$ are given as
\begin{footnotesize}
\begin{eqnarray}
  a_i - a^0_i &=& \sum_k v_{ik}s_k z_k,
\end{eqnarray}
\end{footnotesize}

\noindent
where the scaling factors $\{s_i\}$ are introduced to
normalize $\{z_i\}$. Usually, $\{s_i\}$ are chosen to make

\begin{footnotesize}
\begin{eqnarray}
  \Delta \chi^2 = \chi^2 - \chi^2_\text{min} = \sum_i z^2_i,
\end{eqnarray}
\end{footnotesize}

\noindent so that the best fit of $\chi^2$ (denoted as $\chi^2_\text{min}$)
corresponds to $\{z_i=0\}$, and the one-standard deviation
uncertainty in the direction of the $i$-th eigenvector
corresponds to $\{z_i = \pm 1;z_j=0$ for $j\ne i\}$.
By this requirement, $\{s_i\}$ are approximately equal to $\sqrt{2/\epsilon_i}$. 

In practice, $\chi^2$ is more complicated, which may include 
correlated systematical uncertainties. Furthermore, the one-standard deviation 
uncertainty is usually enlarged by some dynamical and/or global tolerances to 
deal with the possible inconsistency between data, and to cover the uncertainty 
due to the choice of the non-perturbative parameterizations. 
For simplicity, we ignore all these variants in this work as they will 
not strongly alter our conclusions. 

For a given physical observable whose theory prediction depends on PDFs, 
it can be expanded as:

\begin{footnotesize}
\begin{eqnarray}\label{eq:expansion}
 \mathcal{O}(\{z_i\}) = \mathcal{O}(\{z^0_i\}) + \sum_i \frac{\partial \mathcal{O}}{\partial z_i}\Big|_{z^0_i}   \text{d} z_i
    + \frac{1}{2}\sum_{i,j}\frac{\partial^2 \mathcal{O}}{\partial z_i \partial z_j} \Big|_{z^0_i, z^0_j} \text{d}z_i \text{d} z_j + \cdot\cdot\cdot
\end{eqnarray}
\end{footnotesize}

\noindent As $\{z_i\}$ represent orthogonal eigenvectors, 
the leading order
uncertainty on $\mathcal{O}$ can
simply be calculated as:

\begin{footnotesize}
\begin{eqnarray}\label{eq:unc}
 \delta \mathcal{O} &\approx & \sqrt{\sum_i \left[ \frac{\mathcal{O}(z^0_i + \delta z_i) - \mathcal{O}(z^0_i - \delta z_i)}{2\delta z_i} \right]^2}
%   &\approx& \sqrt{\sum_i \left( \frac{1}{2}\frac{\partial \mathcal{O}}{\partial z_i}\Big|_0 \delta z_i   \right)^2}
\end{eqnarray}
\end{footnotesize}

\noindent where $\delta z_i=1$ is the uncertainty on $z_i$.
$\mathcal{O}(z^0_i \pm \delta z_i)$ is the calculation of $\mathcal{O}$
where the value of the $i$-th parameter is shifted by $\pm \delta z_i$, 
while all other parameters $z_j(j\ne i)$
are fixed at their central values $z^0_j$. 
To enable the calculation, the global analysis provides a central PDF set $S_0 = S(\{z^0_i\})$ which
can be used to calculate the central prediction of $\mathcal{O}(S_0)$, and a group of error PDF sets
$S_{i\pm} = S(z^0_i \pm \delta z_i; z_j=z^0_j)$ for $j\ne i$. 
Hence, the uncertainty in Eq.~\eqref{eq:unc} can be calculated using the PDF error sets:

\begin{footnotesize}
\begin{eqnarray}
  \delta \mathcal{O} = \sqrt{\sum_i  \left(  \frac{\mathcal{O}(S_{i+}) - \mathcal{O}(S_{i-})}{2} \right)^2   }
\end{eqnarray}
\end{footnotesize}
~\\

\section{III. Non-linear effects in global analysis}

The first linear assumption in the original Hessian method is introduced in the determination 
of the eigenvectors and their uncertainties. The orthogonal parameters $\{z_i\}$ are acquired using 
the Hessian matrix containing only the $\partial^2 \chi^2 / \partial a_i \partial a_j$ terms. 
The one standard deviation 
is defined as $\delta z_i=1$ corresponding to $\Delta \chi^2=1$. All these settings are based on the 
assumption that the physical observables of the data used in the global analysis can be written as 
the linear functions of the original PDF parameters $\{a_i\}$:

\begin{footnotesize}
\begin{eqnarray}\label{eq:firstLinear}
  T_\alpha (\{a_i\}) = \sum_i t^i_\alpha a_i
\end{eqnarray}
\end{footnotesize}

\noindent where $t^i_\alpha$ is the coefficient. 
In principle, Eq.~\eqref{eq:firstLinear} never holds true in global analysis.
Firstly, the non-perturbative functions 
in Eq.~\eqref{eq:npfunction} are 
naturally non-linear, e.g. the power terms $x^{1-a_1}$ and $(1-x)^{a_2}$ which 
are guided by the Regge theory~\cite{Regge} and the spectator counting rules~\cite{SPCount} to 
describe the asymptotic behavior of $f_q(x)$ in the limits $x\rightarrow 0$ and 1. 
Secondly, the calculations on physical interactions are usually 
convolutions over $x$. 
Thirdly, for those collider and fixed target experiments, the calculations 
on the two-hadron initial state contain the product of the two parton densities. 
In practice, when data inputs used in a global analysis have high precisions, 
$\Delta \chi^2$ would increase rapidly as $\{a_i\}$ 
values running away from their best fitted values. Thus, 
all calculations can be done in small intervals 
around $\{a^0_i\}$ so that $T_\alpha(\{a_i\})$ can be numerically linear. 
In other words, the non-linearity depends on the precision of the 
experimental measurements in the PDF global analysis.
In current PDFs, the 
experimental inputs provide good constraints in some $x$ regions and 
for the quark flavors that they are sensitive to. However, 
the data precisions in the foreseeable future would not be good enough 
to suppress the non-linear effect 
for all quark flavors and in a wide range of $x$.

Another linear assumption is introduced in computing the PDF-induced 
uncertainties on physical observables using the eigenvectors. 
In the current Hessian method, the higher order contributions in 
Eq.~\eqref{eq:expansion} are 
ignored when extrapolating the uncertainties on eigenvectors to a given physical observable. 
The Hessian-type PDFs such as CT18 and MSHT20 
provide only the error sets $S_{i\pm}$ corresponding to the complete differential of $z_i$, which 
allows calculation of $\partial \mathcal{O}/\partial z_i$, $\partial^2 \mathcal{O}/\partial z^2_i$ or 
even higher order contributions as the diagonal elements. But 
these error sets cannot give estimations on 
the non-linear terms such as $\partial^2 \mathcal{O}/\partial z_i\partial z_j$. 
The higher order contributions can be significant especially when the computed 
physical observable corresponds to $x$ region or quark flavor very differently from 
that of the data in the global fit. 

In CT18~\cite{CT18}, the uncertainties of the PDF error sets are 
quoted to cover the difference from varying the 
formalism of the non-perturbative functions of Eq.~\eqref{eq:npfunction}. 
By doing this, CT18 gives more conservative uncertainty estimation 
compared to MSHT20~\cite{MSHT20} and NNPDF4~\cite{NNPDF4}, which 
use similar data set in the global fit.
The enlarged uncertainties reflect the non-linear effects 
in global analysis, because   
if the non-perturbative functions tested by CT18 were all in good linear 
approximation, they would lead to consistent predictions on 
any physical observable because they were connected by 
linear transformations in the PDF parameter space. 
In other words, it is, to some extent, similar to estimating the uncertainty due to the non-linearity 
with respect to the uncertainty due to the particular choice of non-perturbative functions used in 
the global fit. 
However, it is inevitable in the enumeration method that many functions being tested 
are known to have disadvantages in describing the data. Consequently, the uncertainty 
may be overestimated by quoting the imperfection of the excluded functions 
rather than the imperfection of the chosen function. On the other hand, the uncertainty 
may also be underestimated as there is always possible functions missed in the 
enumeration. 
Therefore, a new method to estimate the uncertainty of the chosen function itself would be 
important. 

\section{IV. Improved Hessian method}

The improved method will be derived in this Section. Then, in Section V we will 
apply it to a pseudo analysis, to test how non-linear contributions 
can affect a global analysis. The pseudo-linear calculation will be compared to 
the nominal (un-improved) global fit.

\begin{figure}[!h]
\begin{center}
\epsfig{scale=0.4, file=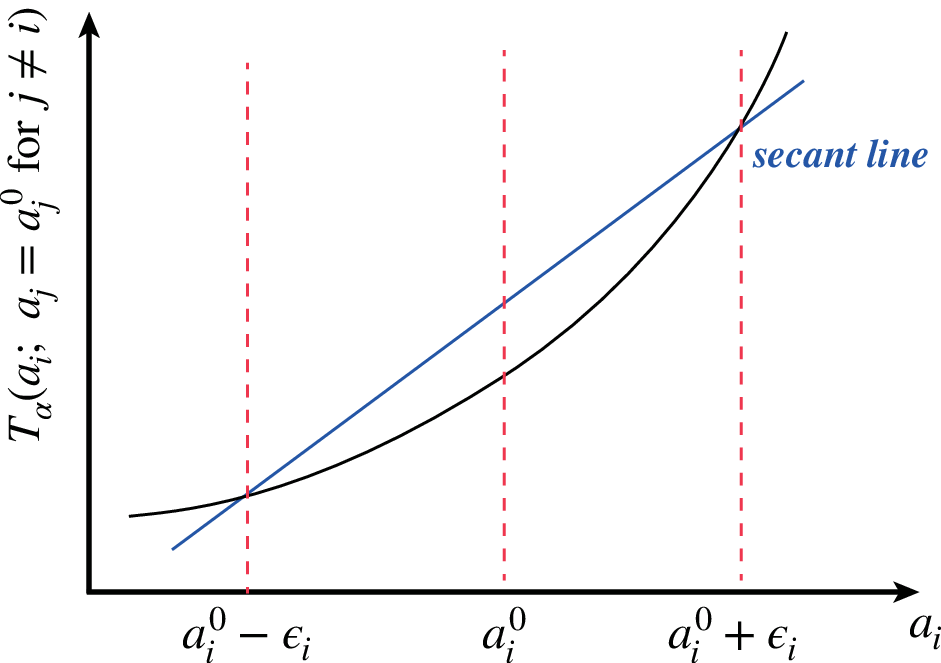}
\caption{\small The secant line of $T_\alpha(a_i;a_j=a^0_j\text{ for }j\ne i)$ around the best fitted value $a^0_i$.}
\label{fig:secant}
\end{center}
\end{figure}

The strategy is as following:

i) For a given non-perturbative function formalism such as in Eq.~\eqref{eq:npfunction}, 
perform the global analysis with the normal method, giving the best fitted values of $\{a_i=a^0_i\}$ and 
the central PDF set $S_0$. In this nominal fit, $T_\alpha(\{a_i\})$ are predicted with different 
$\{a_i\}$ inputs based on the conventional calculations. 

ii) For each physical observable, a set of linear factors are defined as:

\begin{footnotesize}
\begin{eqnarray}\label{eq:secantline}
  K^\alpha_i &=& \frac{T^{i+}_\alpha - T^{i-}_\alpha}{2\epsilon_i}.
\end{eqnarray}
\end{footnotesize}

\noindent where 
$T^{i+}_\alpha = T_\alpha(\{a_i=a^0_i+\epsilon_i;a_j=a^0_j\text{ for }i\ne j\})$ and 
$T^{i-}_\alpha = T_\alpha(\{a_i=a^0_i-\epsilon_i;a_j=a^0_j\text{ for }i\ne j\})$. 
Apparently, $K^\alpha_i$ is an approximation of $\partial T_\alpha(\{a_i\})/\partial a_i$ around 
$\{a_i = a^0_i\}$ in the range of $(a_i-\epsilon_i, a_i+\epsilon_i)$. 
It represents the slope of the secant line, as shown in Fig.~\ref{fig:secant}. 
The $\epsilon_i$ parameter corresponds to the range of $a_i$ in which 
the linear derivation of 
Eq.~\eqref{eq:secantline} is assumed. Thus, $\epsilon_i$ 
can simply be set to the original uncertainty of the $a_i$ parameter given by the fit obtained form i).

iii) A group of new observables can then be established as linear functions of $\{a_i\}$:

\begin{footnotesize}
\begin{eqnarray}\label{eq:lineartheory}
  T^{(r)}_\alpha(\{a_i\}) = T_\alpha(\{a^r_i\}) + \sum_i K^\alpha_i \times (a_i - a^r_i),
\end{eqnarray}
\end{footnotesize} 

\noindent where $\{a^r_i\}$, $r=0, 1, \cdot\cdot\cdot, n$ represent a set of points in the 
parameter space of $\{a_i\}$, with $n$ as the number of the parameters. 
Values of $\{a^r_i\}$ also represent the range of the linear assumption, but  
in terms of $T_\alpha$. In order to achieve this, 
$\{a^r_i\}$ values can be set corresponding to $\{z_r = \Delta_r; z_i = 0\text{ for }i\ne r\}$. 
$\Delta_r$ is chosen to have $T_\alpha(\{a^r_i\}) - T_\alpha(\{a^0_i\})$ 
consistent with $\sigma_\alpha$ for those data inputs sensitive to $z_r$.  Note that 
when the eigenvector $z_r$ is constrained by multiple data inputs in the global fit, 
the uncertainty of $z_r$ ($\delta z_r$) will be smaller than $\Delta_r$, since 
$\delta z_r$ reflects the combination of the statistical power of those data inputs.

Using Eq.~\eqref{eq:lineartheory}, one can perform 
a linear fit, in which the best values of $\{a_i\}$ are determined by 
requiring the minimum of $\chi^2(T^{(r)})$ defined as:

\begin{footnotesize}
\begin{eqnarray}\label{eq:newchi2}
 \chi^2(T^{(r)}) = \sum_\alpha \frac{[D_\alpha - T^{(r)}_\alpha(\{a_i\})]^2}{\sigma^2_\alpha}.
\end{eqnarray}
\end{footnotesize}

\noindent  When Eq.~\eqref{eq:firstLinear} is in good approximation, the 
linear fit using Eq.~\eqref{eq:newchi2} would give best fit values, denoted 
as $\{a^{r,0}_i\}$ for each $r$, consistent 
with the nominal fit $\{a^0_i\}$.
Therefore, the difference between the two fits reflects 
the non-linearity of the physical observables $T_\alpha(\{a_i\})$, and can be 
quoted as additional uncertainties. To do this, an additional PDF error set 
is defined as $S_E(r) = S(\{a^{r,0}_i\})$ for each eigenvector. For a 
given physical observable $\mathcal{O}$, the non-linear uncertainty is 
defined as:

\begin{footnotesize}
\begin{eqnarray}\label{eq:unc1}
  \delta \mathcal{O}_E = \sqrt{\sum_r \left[ \mathcal{O}(S_E(r)) - \mathcal{O}(S_0) \right]^2 }
\end{eqnarray}
\end{footnotesize}

It should be emphasized that Eq.~\eqref{eq:unc1} quantifies the impact on non-linearity when extrapolated to a 
physical observable. Such non-linearity introduces a bias in the analysis, which may in turn contribute to the 
PDF uncertainty. The notions of ``bias'' and ``uncertainty'' are not strictly equivalent, although they 
are likely to be of comparable size. A more detailed analysis would be required to establish their precise 
relationship in a specific case.
$\delta \mathcal{O}_E$ is the uncertainty of assuming $T_\alpha(\{a_i\})$ as linear functions, which is 
independent of how to extrapolate the original experimental uncertainties $\sigma_\alpha$ to 
each eigenvector, and further to a physical observable $\mathcal{O}$. As discussed in the 
section II, the error sets $S_{i\pm}$ allow calculations only on the first-order 
derivatives in Eq.~\eqref{eq:expansion}. 
Below, we show how to obtain the PDF error sets for calculating the
full second-order derivatives.
Following the definition of the derivative, the needed error sets are:

\begin{footnotesize}
\begin{eqnarray}\label{eq:NLOset}
  S^{i\pm}_{j\pm} = S(z_i=\pm 1, z_j=\pm 1; z_k=0\ \mathrm{for}\ k\ne i, j),
\end{eqnarray}
\end{footnotesize}

\noindent in which two displacements are varied simultaneously, so that the second-order derivatives 
can be calculated as:

\begin{footnotesize}
\begin{eqnarray}\label{eq:partials}
 \frac{\partial^2 \mathcal{O}}{\partial z_i \partial z_j} &=& \nonumber \\
   \frac{1}{2} &\times& \left( \frac{\mathcal{O}(S^{i+}_{j+}) - \mathcal{O}(S^{i-}_{j+})}{2}
   -  \frac{\mathcal{O}(S^{i+}_{j-}) - \mathcal{O}(S^{i-}_{j-})}{2} \right)
 \end{eqnarray}
\end{footnotesize}

For $i=j$, the second-order error sets can be calculated as $S_{i++} = S(z_i = 2; z_k=0$ for $k\ne i)$ and
$S_{i--} = S(z_i = -2; z_k=0$ for $k\ne i)$. Consequently, we have

\begin{footnotesize}
\begin{eqnarray}
 \frac{\partial^2 \mathcal{O}}{\partial z^2_i} &=& \nonumber \\
     \frac{1}{2} &\times& \left( \frac{\mathcal{O}(S_{i++}) - \mathcal{O}(S_0)}{2}
   -  \frac{\mathcal{O}(S_0) - \mathcal{O}(S_{i--})}{2} \right)
 \end{eqnarray}
\end{footnotesize}

According to Eq.~\eqref{eq:Hmatrix}, the values of the original parameters $\{a_i\}$ corresponding to
$\{z_i=\pm 1,z_j=\pm 1; z_k=0$ for $k\ne i, j\}$ can be calculated, and thus the
second-order error sets $S^{i\pm}_{j\pm}$ can be directly generated. Since the Hessian matrix comes from
the derivatives of $\chi^2$, the second-order error sets can only be acquired together with
the first-order error sets in a complete PDF global analysis. 
With the full second-order derivatives included, the uncertainty on $\mathcal{O}$ is written as:

\begin{footnotesize}
\begin{eqnarray}\label{eq:uncNLO}
  \left( \delta \mathcal{O} \right)^2 = \sum_i \left( \frac{\partial \mathcal{O}}{\partial z_i}\right)^2
     + \frac{1}{2}\sum_i \left( \frac{\partial^2 \mathcal{O}}{\partial z^2_i} \right)^2 + \frac{1}{2}\sum_{i\ne j}\left( \frac{\partial^2 \mathcal{O}}{\partial z_i \partial z_j} \right)^2,
\end{eqnarray}
\end{footnotesize}

\noindent where the first, second and third terms correspond to the contributions from the first-order derivatives,
the second-order diagonal derivatives, and the second-order off-diagonal derivatives
in Eq.~\eqref{eq:expansion}, respectively. The process of proof of Eq.~\eqref{eq:uncNLO} is given in the Appendix. 

One can also
calculate the uncertainty in an alternative way:
~\\
i) give each displacement $z_i$ (along each eigenvector $v_i$)
a random value according to the normalized gaussian distribution $G(0,1)$.
~\\
ii) calculate $\mathcal{O} = \mathcal{O}(S_0) + \sum_i \frac{\partial \mathcal{O}}{\partial z_i} z_i +\frac{1}{2}\sum_{i,j} \frac{\partial^2 \mathcal{O}}{\partial z_i \partial z_j} z_i z_j$, where the derivatives are calculated using the
error sets.
~\\
iii) repeat step i) and ii) many times, and interpret
the many resulting values of $\mathcal{O}$ as a statistical distribution
of $\mathcal{O}$, with its one standard deviation representing the PDF-induced uncertainty of the observable $\mathcal{O}$.
~\\
\noindent Since the first- and second-order derivatives ({\it i.e.} $\frac{\partial \mathcal{O}}{\partial z_i}$ and
$\frac{\partial^2 \mathcal{O}}{\partial z_i\partial z_j}$) can be readily calculated,
the above bootstrap procedure can be quickly done.

The bootstrap procedure gives consistent uncertainty with Eq.~\eqref{eq:uncNLO}. 
It is also convenient for calculating the correlation of the PDF-induced uncertainties between 
two physical observables $\mathcal{O}_1$ and $\mathcal{O}_2$. One can calculate $(\delta\mathcal{O}_1)^2$, 
$(\delta\mathcal{O}_2)^2$ and $(\delta(\mathcal{O}_1+\mathcal{O}_2))^2$ via either aforementioned method. 
The correlation is then easily acquired on the basis of its original definition of correlation:

\begin{footnotesize}
\begin{eqnarray}
  \rho(\mathcal{O}_1,\mathcal{O}_2) = \frac{(\delta(\mathcal{O}_1+\mathcal{O}_2))^2-(\delta\mathcal{O}_1)^2-(\delta\mathcal{O}_2)^2}{2\delta\mathcal{O}_1\delta\mathcal{O}_2}
\end{eqnarray}
\end{footnotesize}

\noindent Note that both the bootstrap method and Eq.~\eqref{eq:uncNLO} assume $\{ z_i \}$ have 
independent gaussian distributions. 
As will be discussed in the next section, it is equivalent to the linear 
assumptions of Eq.~\eqref{eq:firstLinear}, of which the uncertainty has  been estimated as $\delta \mathcal{O}_E$.
In conclusion, the total uncertainty on a physical observable $\mathcal{O}$ is the quadratic combination 
of the original first-order uncertainties, the second-order uncertainties, and the non-linear uncertainty of 
$\delta \mathcal{O}_E$:

\begin{footnotesize}
\begin{eqnarray}\label{eq:uncTotal}
 && \left( \delta \mathcal{O}_\text{total} \right)^2 = \frac{1}{4}\sum_i \left[ \mathcal{O}(S_{i+})-\mathcal{O}(S_{i-})\right]^2  \nonumber \\
   &+&  \frac{1}{32} \sum_i \left[  \mathcal{O}(S_{i++})+\mathcal{O}(S_{i--})-2\mathcal{O}(S_0) \right]^2  \nonumber \\
     &+& \frac{1}{32} \sum_{i\ne j} \left[ \mathcal{O}(S^{i+}_{j+})-\mathcal{O}(S^{i-}_{j+}) - 
           \mathcal{O}(S^{i+}_{j-})  + \mathcal{O}(S^{i-}_{j-}) \right]^2 \nonumber \\
     &+& \sum_r \left[ \mathcal{O}(S_E(r)) - \mathcal{O}(S_0) \right]^2 .
\end{eqnarray}
\end{footnotesize}

The various contributions to the uncertainty can be numerically calculated with different settings. For example, 
in Eq.~\eqref{eq:NLOset}, we vary $z_i$ and $z_j$ from 0 to $\pm 1$. 
Actually the variation can be any value as long as the denominator in Eq.~\eqref{eq:partials} changes 
accordingly. Similarly, the calculations of second-order derivatives in Eq.~\eqref{eq:partials} are 
not exclusive. 
Again, the non-linear uncertainty depends on the choice of 
$\epsilon_i$ and $\{a^r_i\}$. 
In a real global analysis, these settings should be determined with multiple factors taken into consideration, 
such as the particular data observables used 
in the global analysis, the experimental precision, 
the overall fitting quality, and the potential tensions between data. 
It also depends on the confidence level of a specific global 
analysis. For example, CTEQ-TEA gives its original uncertainty at 90$\%$ C.L.~\cite{CT18}, while 
NNPDF directly estimates its uncertainty at 68.3$\%$ C.L.~\cite{NNPDF4}. 
In this study we have focused on describing a general method instead of 
discussing the detailed choice of numerical settings, which should be decided 
in real global analysis on a case by case basis. 

~\\

\section{V. Pseudo global analysis}

\begin{figure}[!h]
\begin{center}
\epsfig{scale=0.4, file=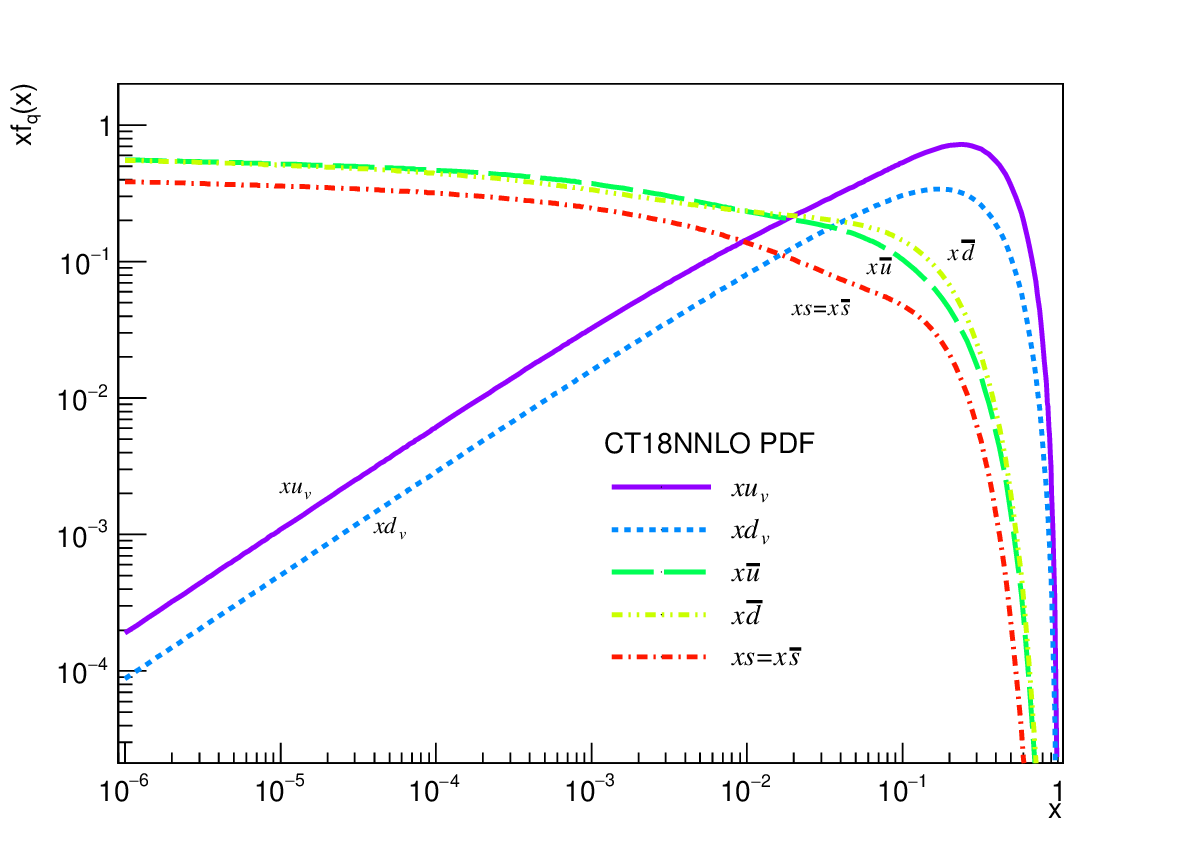}
\caption{\small $xf_q(x)$ for $u_V$, $d_V$, $\bar{u}$, $\bar{d}$ and $s=\bar{s}$ in the pseudo
global analysis.}
\label{fig:PDFs}
\end{center}
\end{figure}

In this section, we introduce a pseudo global analysis to numerically test the improved Hessian method.
We start with
a set of parton densities taken from the CT18NNLO PDFs~\cite{CT18}. For valence $u_V$ and $d_V$,
the non-perturbative functions are:

\begin{footnotesize}
\begin{eqnarray}\label{eq:valence}
  f_q(x) &=& a_0 x^{a_1 - 1}(1-x)^{a_2}P^V_a(y) \nonumber \\
    P^V_a(y) &=& \sinh[a_3](1-y)^4 + \sinh[a_4]4y(1-y)^3 + \sinh[a_5]6y^2(1-y)^2 \nonumber \\
      && + \left(1+\frac{1}{2}a_1 \right) 4y^3(1-y) + y^4,
\end{eqnarray}
\end{footnotesize}

\noindent where $y\equiv \sqrt{x}$. For both $u_V$ and $d_V$, $a_0$ is
fixed, following the flavor sum rules. We further require $a^{u_V}_1 = a^{d_V}_1$ and
$a^{u_V}_2 = a^{d_V}_2$, as done in CT18 global analysis.
For the sea quarks $\bar{u}$, $\bar{d}$, and $s=\bar{s}$, the
non-perturbative functions are:

\begin{footnotesize}
\begin{eqnarray}\label{eq:sea}
  f_q(x) &=& a_0 x^{a_1 - 1}(1-x)^{a_2}P^{\text{sea}}_a(y) \nonumber \\
    P^{\text{sea}}_a(y) &=&  (1-y)^5 + a_4 5y(1-y)^4 + a_5 10y^2(1-y)^3 \nonumber \\
    & & + a_6 10 y^3 (1-y)^2 + a_7 5y^4(1-y) + a_8y^5,
\end{eqnarray}
\end{footnotesize}

\noindent where $y\equiv 1- (1-\sqrt{x})^{a_3}$. We require $a^{\bar{u}}_0 = a^{\bar{d}}_0$,
$a^{\bar{u}}_1 = a^{\bar{d}}_1 = a^{s}_1$, $a^{\bar{u}}_2 = a^{\bar{d}}_2$,
$a^{\bar{u}}_3 = a^{\bar{d}}_3 = a^{s}_3 = 4$, $a^{\bar{d}}_8=a^{s}_8 = 1$,
$a^{s}_4=a^s_5$, and $a^s_6=a^s_7$, following CT18. The distribution of the gluon is ignored in this test,
since it would complicate the PDF analysis to add gluon contributions in the pseudo data.
The charm quark is also ignored because it is highly correlated with the gluon.
Therefore, we have 24 free parameters in total. Their central values are also
set to the best fit of CT18NNLO at the initial energy scale $Q_0=1.3$ GeV~\cite{CT18}, as listed in Tab.~\ref{tab:inputPar}.
Figure~\ref{fig:PDFs} depicts $xf_q(x)$ for different quarks. We note that in this study, these pseudo PDFs will not be evolved to a high energy scale relevant to the pseudo data introduced below. These pseudo PDFs do not resemble any of the CT18 PDF sets. They are merely designed to simplify our discussions on the effect of the second-order derivatives in Eq.~\eqref{eq:expansion} on the estimates of PDF-induced uncertainties.

\begin{table}[hbt]
\begin{footnotesize}
\begin{center}
\begin{tabular}{l|c|c|c|c|c}
\hline \hline
   & $u_V$ & $d_V$ & $\bar{u}$ & $\bar{d}$ & $s=\bar{s}$ \\
 \hline
 $a_0$ & 3.385 (fixed) & 0.490 (fixed) & 0.414 & 0.414 & 0.288 \\
\hline
 $a_1$ & 0.763 & 0.763 & -0.022 & -0.022 & -0.022 \\
\hline
 $a_2$ & 3.036 & 3.036 & 7.737 & 7.737 & 10.31 \\
\hline
 $a_3$ & 1.502 & 2.615 & 4 (fixed) & 4 (fixed) & 4(fixed) \\
\hline
 $a_4$ & -0.147 & 1.828 & 0.618 & 0.292 & 0.466 \\
\hline
 $a_5$ & 1.671 & 2.721 & 0.195 & 0.647 & 0.466 \\
\hline
 $a_6$ & - & - & 0.871 & 0.474 & 0.225 \\
\hline
 $a_7$ & - & - & 0.267 & 0.741 & 0.225 \\
\hline
 $a_8$ & - & - & 0.733 & 1 (fixed) & 1 (fixed) \\
 \hline \hline
\end{tabular}
\caption{\small Values of parameters in the non-perturbative functions, taken from CT18NNLO.}
\label{tab:inputPar}
\end{center}
\end{footnotesize}
\end{table}

A group of pseudo observables are also constructed according to the PDFs in
Eq.~\eqref{eq:valence} and Eq.~\eqref{eq:sea}. These observables, with uncertainties
assigned, will be used as the data input in the pseudo global analysis.
First, we introduce $\sigma^\text{DIS}_\gamma(x)$ and $\sigma^\text{DIS}_Z(x)$:

\begin{footnotesize}
\begin{eqnarray}
 \sigma^\text{DIS}_\gamma (x) &=& u_V(x) + 2\bar{u}(x) + 0.25\left[ d_V(x) + 2\bar{d}(x) + 2s(x)\right] \nonumber \\
 \sigma^\text{DIS}_Z(x) &=& u_V(x) + 2\bar{u}(x) + 1.2\left[ d_V(x) + 2\bar{d}(x) + 2s(x)\right]
\end{eqnarray}
\end{footnotesize}

\noindent as pseudo observables similar to the cross section measurements
of the photon-exchange DIS process $\ell + p \rightarrow \gamma^* \rightarrow \ell+X$, and
the $Z$-exchange DIS process $\ell + p \rightarrow Z/\gamma^* \rightarrow \ell+X$. The $q(x)$ functions
are taken from Eq.~\eqref{eq:valence} and Eq.~\eqref{eq:sea}, with
$\{a_i\}$ values from Tab.~\ref{tab:inputPar}.
The coefficient 0.25 (1.2) roughly reflects
the strength difference between the down-type and up-type quarks coupling to $\gamma$ ($Z$).
Note that we take the definitions of $u \equiv u_V + \bar{u}$ and $d \equiv d_V + \bar{d}$,
so that the $\bar{u}$ and $\bar{d}$ terms have a coefficient of 2.
For simplicity, we ignore other contributions
in a real calculation of the DIS process, such as the evolution of
PDFs as a function of $Q$, the electroweak couplings and
the higher order corrections. Although these effects would significantly
change the predictions on any physical observable, the main sensitivity to
the particular parton information in those observables remains.
For $\sigma^\text{DIS}_\gamma(x)$, 28 data points are calculated,
with $x =$ 0.0001, 0.0002, 0.0003, 0.0004, 0.0005, 0.0006, 0.0007,
0.0008, 0.0009, 0.001, 0.002, 0.003, 0.004, 0.005, 0.006, 0.007, 0.008, 0.009, 0.01, 0.02, 0.03, 0.04, 0.05, 0.06,
0.07, 0.08, 0.09, 0.1. For $\sigma^\text{DIS}_Z(x)$, 16 points are calculated, with
$x = $ 0.01, 0.02, 0.03, 0.04, 0.05, 0.06, 0.07, 0.08, 0.09, 0.1, 0.15, 0.2, 0.25, 0.3, 0.35, 0.4.

Similarly, we introduce the following cross section observables in 
proton-proton collisions: $\sigma^\text{pp}_Z(x_1, x_2)$, $\sigma^\text{pp}_{W^+}(x_1, x_2)$ and
$\sigma^\text{pp}_{W^-}(x_1, x_2)$ as:

\begin{footnotesize}
\begin{eqnarray}
 \sigma^\text{pp}_Z(x_1, x_2) &=& \left[ u_V(x_1) + \bar{u}(x_1) \right]\bar{u}(x_2) +  \left[ u_V(x_2) + \bar{u}(x_2) \right]\bar{u}(x_1) \nonumber \\
     & & + 1.2 \left[ d_V(x_1) + \bar{d}(x_1) \right]\bar{d}(x_2)   \nonumber \\
     && + 1.2 \left[ d_V(x_2) + \bar{d}(x_2) \right]\bar{d}(x_1) \nonumber \\
     && + 1.2 \times 2 \times s(x_1)s(x_2) \nonumber \\
 \sigma^\text{pp}_{W^+}(x_1, x_2) &=& 0.9 \left[ (u_V(x_1)+\bar{u}(x_1))\bar{d}(x_2) \right]  \nonumber \\
       & & + 0.9 \left[ (u_V(x_2)+\bar{u}(x_2))\bar{d}(x_1) \right]  \nonumber \\
       & & + 0.1 \left[ (u_V(x_1)+\bar{u}(x_1))s(x_2) \right] \nonumber \\
       & & + 0.1 \left[ (u_V(x_2)+\bar{u}(x_2))s(x_1) \right] \nonumber \\
 \sigma^\text{pp}_{W^-}(x_1, x_2) &=& 0.9 \left[ (d_V(x_1)+\bar{d}(x_1))\bar{u}(x_2) \right]  \nonumber \\
       & & + 0.9 \left[ (d_V(x_2)+\bar{d}(x_2))\bar{u}(x_1) \right]  \nonumber \\
       & & + 0.1 \bar{u}(x_1)s(x_2) + 0.1 \bar{u}(x_2)s(x_1)
\end{eqnarray}
\end{footnotesize}

\noindent They approximately represent the vector boson productions of
$pp(q\bar{q})\rightarrow Z/\gamma^* \rightarrow \ell^+\ell^-$ and
$pp(q_i\bar{q}_j)\rightarrow W^{\pm} \rightarrow \ell^\pm +\nu$. The momentum fractions
$x_1$ and $x_2$ of the two initial quarks
are connected by
$$
x_{1,2} = \frac{\sqrt{M^2 + Q^2_T}}{\sqrt{s}} e^{\pm Y}
$$
where $M$, $Q_T$ and $Y$ are the invariant mass, transverse momentum and rapidity of the bosons, respectively.
In this work, $M$ is set to 90 GeV, which is near the pole of the $W$ and $Z$ boson masses.
Compared to $M$, the transverse momentum of the single vector boson production
is usually small. Therefore $Q_T$ is ignored in this work. $\sqrt{s}$ is set to 13 TeV.
To gain information for different $x$ values, the cross section observables are
calculated with different $|Y|$ values of
0.1, 0.3, 0.5, 0.7, 0.9, 1.1, 1.3, 1.5, 1.7, 1.9, 2.1, 2.4, 2.8, 3.2, 4.0.
The corresponding $x$ values range from 0.0001 to 0.38.
 
We also introduce a group of new experimental observables into the global fit, which are the 
proton structure parameters, $P_u$ and $P_d$, factorized from the forward-backward asymmetry of the
Drell-Yan process~\cite{AFBFactorization},
and the boost asymmetry $A_\text{boost}$ in the diboson process~\cite{BoostAsymmetry}.
These asymmetry variables were recently proposed at hadron colliders which can provide
unique information on the proton structure.
The $P_u$ and $P_d$ parameters factorize the proton structure information
in $pp(u\bar{u})\rightarrow Z/\gamma^* \rightarrow \ell^+\ell^-$ and
$pp(d\bar{d})\rightarrow Z/\gamma^* \rightarrow \ell^+\ell^-$ events.
The two light quark initial states can be decoupled because the electroweak
couplings of $u$-$\bar u$-$Z$ and $d$-$\bar d$-$Z$ are different. As a result,
$P_u$ and $P_d$ separately provide the $u$ and $d$ quark information, which
are mixed and indistinguishable in the measurement of the total cross section $\sigma^{pp}_Z$.
$A_\text{boost}$ is the asymmetry defined in the diboson production~\cite{BoostAsymmetry}. For example, in
the $pp(u\bar{d}) \rightarrow W^+\gamma \rightarrow \ell^+\nu+\gamma$ process,
the $W^+$ boson predominantly couples to the $\bar{d}$ quark while the photon couples to the
$u$ quark. Consequently, the boost asymmetry $A_{\rm boost}^{W^+ \gamma}$ has a positive large value, namely $\gamma$ is more boosted than the $W^+$ boson, because the initial state parton lunminosity $u(x_1){\bar d}(x_2)$ is larger than  $u(x_2){\bar d}(x_1)$ for $x_1 > x_2$, etc. Therefore, the relative asymmetry between the
initial-state parton luminosities $u(x_1)\bar{d}(x_2)$ and $u(x_2)\bar{d}(x_1)$ ($x_1>x_2$) 
can be measured by comparing the energy or rapidity of the two bosons.

According to Ref.~\cite{AFBFactorization} and Ref.~\cite{BoostAsymmetry},
the new pseudo observables can be constructed as:

\begin{footnotesize}
\begin{eqnarray}
  P_u(x_1, x_2) &=& \left[ u_V(x_1)+\bar{u}(x_1)\right]\bar{u}(x_2) - \left[ u_V(x_2)+\bar{u}(x_2)\right]\bar{u}(x_1), \nonumber \\
  P_d(x_1, x_2) &=& \left[ d_V(x_1)+\bar{d}(x_1)\right]\bar{d}(x_2) - \left[ d_V(x_2)+\bar{d}(x_2)\right]\bar{d}(x_1), \nonumber \\
  A^{W^+\gamma}_\text{boost}(x_1, x_2) &=&  \left[ u_V(x_1)+\bar{u}(x_1)\right]\bar{d}(x_2) - \left[ u_V(x_2)+\bar{u}(x_2)\right]\bar{d}(x_1), \nonumber \\
  A^{W^-\gamma}_\text{boost}(x_1, x_2) &=&  \left[ d_V(x_1)+\bar{d}(x_1)\right]\bar{u}(x_2) - \left[ d_V(x_2)+\bar{d}(x_2)\right]\bar{u}(x_1). \nonumber\\
  & &
\end{eqnarray}
\end{footnotesize}

\noindent In summary, 149 pseudo data points are introduced, corresponding to the DIS and Drell-Yan
processes that are actually used in the PDF global analysis.
For each data point, we assign a relative uncertainty of $3\%$. This precision is 
comparable to the HERA measurements and the LHC Drell-Yan measurements.
For experimental results other than HERA and LHC, such as earlier DIS and semi-inclusive
DIS measurements, their precisions are much lower. The experiments actually used in
the PDF global analysis are summarized in Ref.~\cite{CT18}.
Although these
pseudo observables are only rough estimations rather than real calculations, they reflect the
complexity of using multiple data inputs to determine the PDF shape parameters.

\begin{figure}[!hbt]
\begin{center}
\epsfig{scale=0.4, file=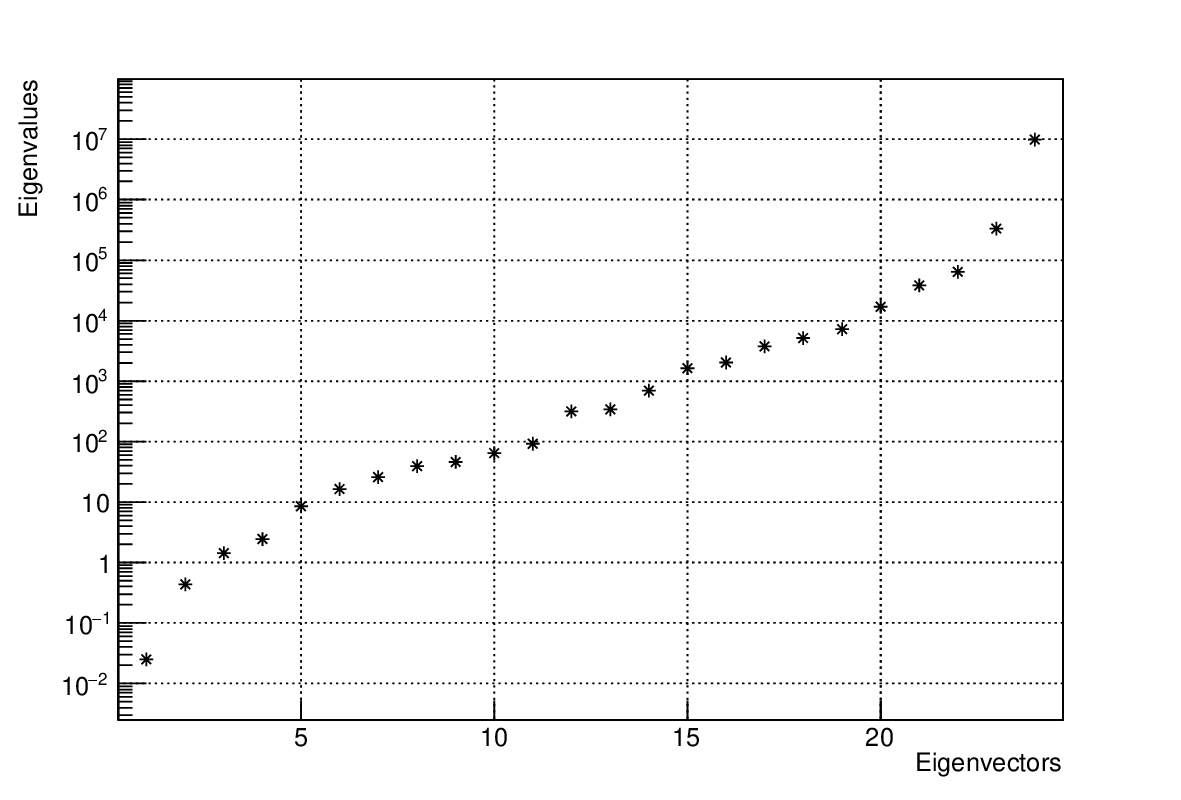}
\caption{\small 24 eigenvalues calculated from the Hessian matrix in the pseudo global analysis.}
\label{fig:eigenvalue}
\end{center}
\end{figure}

With the constructed PDFs, the pseudo observables and the assigned uncertainty, the
Hessian matrix can be calculated accordingly.
Since there are 24 free parameters in this test, the same number
of eigenvectors and corresponding eigenvalues are acquired
by diagonalizing the Hessian matrix.
The eigenvalues in this test are shown in Fig.~\ref{fig:eigenvalue}.
As explained in Ref.~\cite{Hessian}, the eigenvalues
in logarithmic scale rank as a line, reflecting the sensitivities of the input data.
An eigenvector with large eigenvalue usually indicates that the information related to this
eigenvector is well constrained by the input data.
Otherwise, the input data has poor sensitivity to this eigenvector, leading to a small eigenvalue.
Although data points in this analysis are much more
than the free parameters being determined and the assigned uncertainty of $3\%$ is quite small,
there is still PDF information for certain $x$ regions and parton flavors, for which no
pseudo observable gives direct sensitivity.
Therefore, some eigenvalues are small compared to others.
This also happens in a real PDF global analysis~\cite{CT18, MSHT20, HERA}.

\begin{figure}[!hbt]
\begin{center}
\epsfig{scale=0.4, file=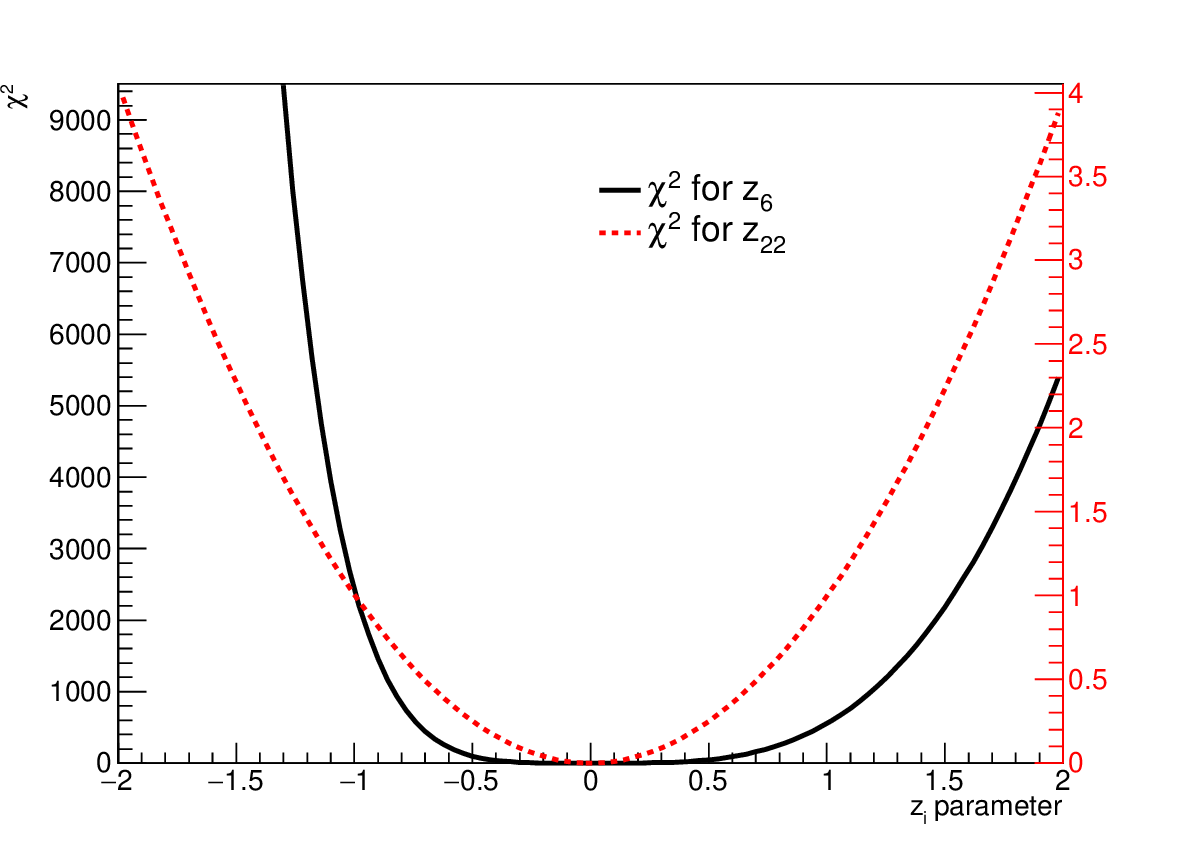}
\caption{\small $\chi^2$-$z_i$ relationship for $i=22$ (demonstrated in the right vertical axis) and
$i=6$ (demonstrated in the left vertical axis) eigenvectors.}
\label{fig:chi2vsz}
\end{center}
\end{figure}

Once the Hessian matrix is diagonalized and the orthogonal displacements
$\{z_i\}$ are acquired, one can look into the $\chi^2$-$z_i$ relationship. 
It is a good numerical test of the linear assumptions in the global analysis.
Usually,
if the eigenvector represented by $z_i$ is well constrained, $\chi^2$ is expected to be
an ideal quadratic function of $z_i$, giving
$\chi^2(z_i=\pm n; z_j=0$ for $j\ne i) = n^2$.  
As discussed before, it also means the data inputs related to this 
eigenvector $z_i$ are precise enough so that the corresponding $T_\alpha$ are 
approximately linear functions of the original parameters $\{a_i\}$ within small intervals. 
In Fig.~\ref{fig:chi2vsz}, the
$\chi^2$-$z_i$ relationship for $i=22$ and $i=6$ are shown and compared, as an example.
$i=22$ is a typical eigenvector which has good linear approximation. Its eigenvalue is 
large at $\mathcal{O}(10^4)$. The $\chi^2$-$z_i$ relationship follows
 a good quadratic approximation, so that $z_i=1$ yields 
$\Delta \chi^2 = \chi^2 - \chi^2_\text{min} = 1$ corresponding to 1 standard deviation. 
For $i=6$ however, it is an eigenvector which is not well constrained by data, 
resulting in a non-symmetric $\chi^2$-$z_i$ relationship. 
In practice (also in this work), the 1 standard deviation in the uncertainty estimation is usually defined 
with the $\{z_i\}$ values which yields $\Delta \chi^2 = \chi^2 - \chi^2_\text{min} = 1$. 

~\\

\begin{figure*}[hbt]
\begin{center}
\epsfig{scale=0.35, file=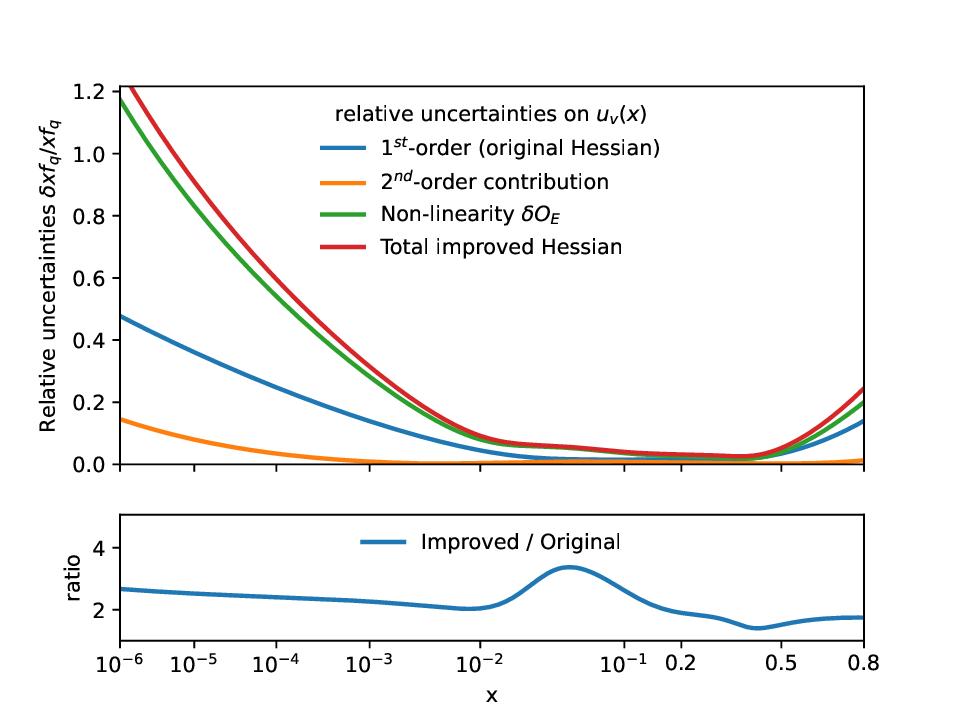}
\epsfig{scale=0.35, file=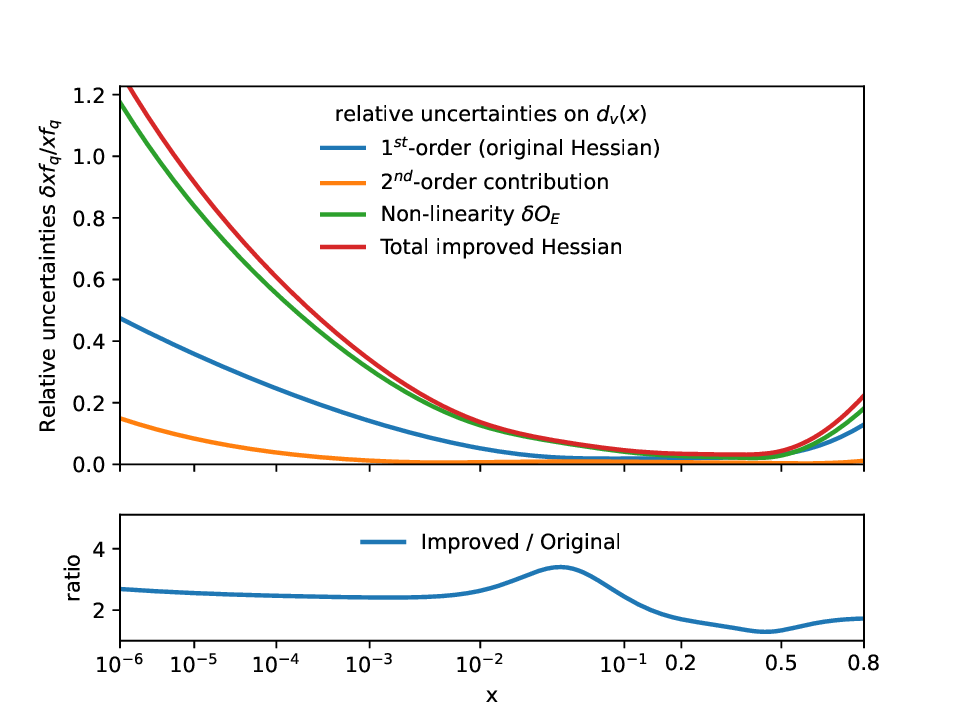}
\epsfig{scale=0.35, file=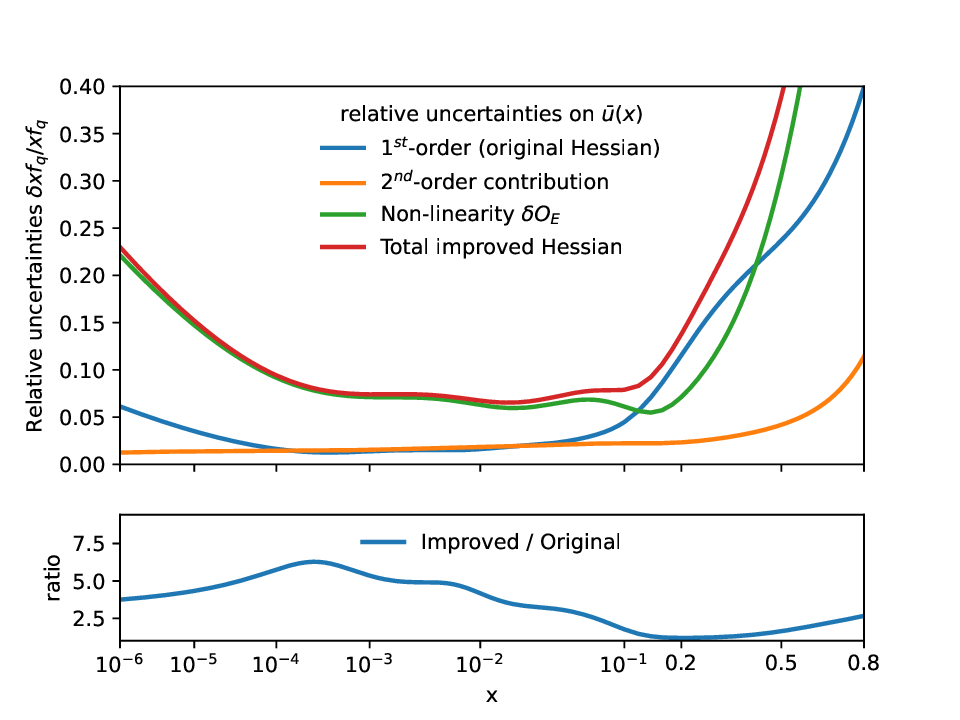}
\epsfig{scale=0.35, file=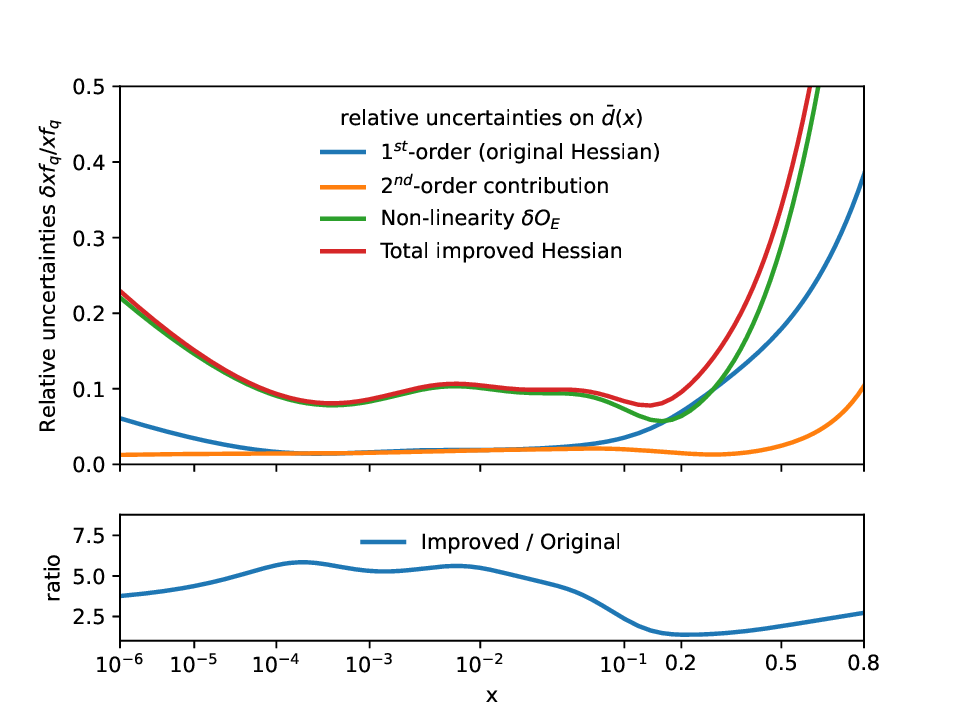}
\epsfig{scale=0.35, file=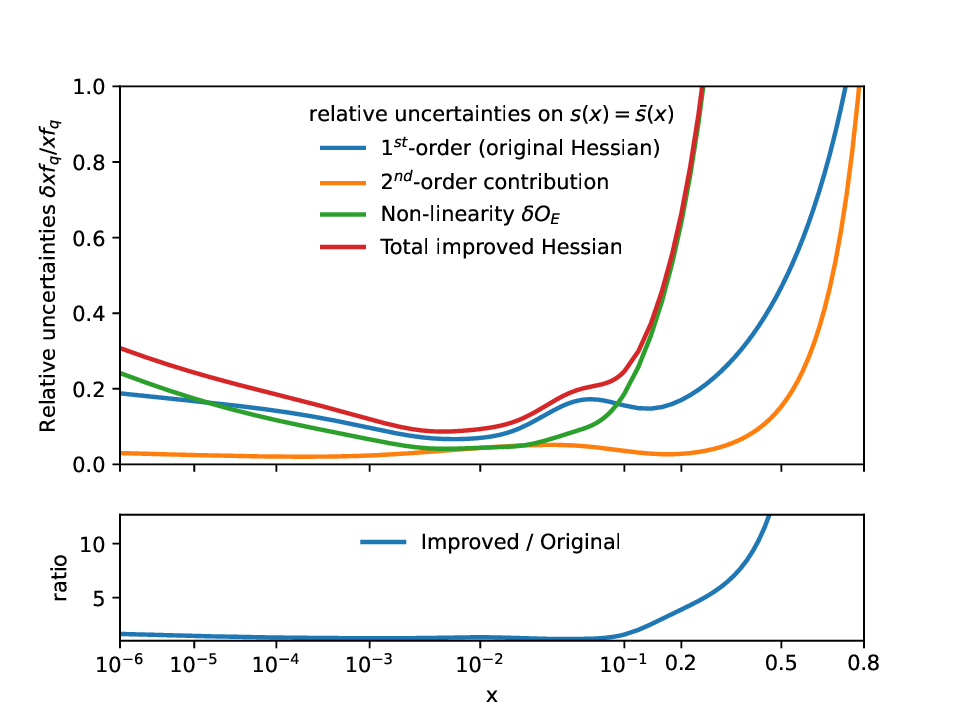}
\caption{\small Relative uncertainties on $xf_{u_V}(x)$, $xf_{d_V}(x)$,
$xf_{\bar{u}}(x)$, $xf_{\bar{d}}(x)$ and $xf_s(x)$. 
Comparison includes the contribution 
of the first-order derivatives which is estimated from the 
original Hessian method (blue curves), the uncertainty of $\delta \mathcal{O}_E$ which estimates 
the non-linearity of $T_\alpha(a_i)$ (green curves), the uncertainty of the second-order derivatives 
additional to the first-order contributions (orange curves), and the total uncertainty from 
the improved Hessian method which is the quadratic sum of the previous three (red curves).
The lower panel in each plot gives the
ratio of the total uncertainty from the improved Hessian method to the original Hessian method.}
\label{fig:results_pdf}
\end{center}
\end{figure*}

Following the method described in section IV, 
the uncertainty of any physical observable $\mathcal{O}$ can be estimated according to Eq.~\eqref{eq:uncTotal}.
To numerically study the difference between the original Hessian method and the improved
Hessian method, we compare the relative uncertainties of $xf_q(x)$ for
$u_V$, $d_V$, $\bar{u}$, $\bar{d}$ and $s$, which are shown
in Fig.~\ref{fig:results_pdf}. 

Fig.~\ref{fig:results_pdf} shows that the non-linear uncertainty 
in this pseudo global analysis is much larger than the original uncertainty, especially 
in the large and small $x$ region where no data can provide direct constraint. 
Even though the pseudo data inputs in this work are generated using the non-perturbative functions 
exactly the same as that used in the pseudo theory calculations, the uncertainty of the choice of 
non-perturbative functions can still be reflected in the non-constrained area. 
The DIS-liked observables are dominated by the valence quarks of $u_V$ and $d_V$ around 
the peak region of their distributions at $x\sim 0.1$, where the contributions of the sea quarks 
are relatively low. For the sea quarks however, there is no data predominated by them. 
Therefore, the valence pseudo quark PDFs are constrained more directly than the sea quark PDFs 
in this study. 
Consequently, the non-linear uncertainties for 
$u_V$ and $d_V$ are generally smaller than that for the sea quarks, as shown in Fig.~\ref{fig:results_pdf}. 
Moreover, the second-order uncertainty on the parton densities is not as significant 
as the non-linear uncertainty. 
If a physical observable contains information from multiple partons, 
the second-order uncertainties could be larger. 
In practice however, 
it could be fine not to compute the second-order uncertainty with thousands of 
second-order error sets for simplicity, especially when the non-linear uncertainty is 
already quoted. 

In this work, the total uncertainties in the 
pseudo global analysis is several times larger than the original ones. 
According to the study from the CTEQ-TEA PDFs, the original uncertainties have to be 
enlarged corresponding to $\Delta \chi^2 = 37$, at the $68\%$ confidence level, to cover the difference of using 
various non-perturbative functions, which also leads to a total uncertainty several times larger. 
The MSHT20 PDF gives a similar conclusion. 
In their studies, the PDF uncertainty is defined by a dynamic tolerance which 
effectively corresponds to $\Delta \chi^2 \sim 10$, giving the final uncertainty 
about 3 times larger than the original one. As a conclusion, the non-linear effect in the 
Hessian method is essential according to today's experimental precisions. 
~\\

\section{Acknowledgement}
\begin{acknowledgements}
This work was supported by the National Natural Science Foundation of China under Grant No. 11721505, 11875245,
12061141005 and 12105275, and supported by the ``USTC Research Funds of the Double First-Class Initiative''.
This work was also supported by the U. S. National Science Foundation under Grant No. PHY-2310291.
~\\
\end{acknowledgements}

\begin{appendix}
\section{Appendix: PDF-induced uncertainty on physical observables}

In this section, we give the detailed calculation of Eq.~\eqref{eq:uncNLO}.
Rewrite Eq.~\eqref{eq:expansion} as:

\begin{footnotesize}
\begin{eqnarray}
  \mathcal{O}(z_i) &=& \mathcal{O}(z^0_i) + \sum_i A_i + \frac{1}{2} \sum_{i\ne j} B_{ij} \nonumber \\
    A_i &=& \frac{\partial \mathcal{O}}{\partial z_i} z_i + \frac{1}{2} \frac{\partial^2 \mathcal{O}}{\partial z^2_i} z^2_i \nonumber \\
    B_{ij} &=& \frac{\partial^2 \mathcal{O}}{\partial z_i \partial z_j} z_i z_j
\end{eqnarray}
\end{footnotesize}

The standard deviation of $A_i$ can be calculated as:

\begin{footnotesize}
\begin{eqnarray}
  \text{V}[A_i] &=& \text{E}[A^2_i] - \text{E}[A_i]^2 \nonumber \\
    &=& \text{E}\left[ \left( \frac{\partial \mathcal{O}}{\partial z_i} \right)^2 z^2_i  \right] + \text{E}\left[  \frac{\partial \mathcal{O}}{\partial z_i}\frac{\partial^2 \mathcal{O}}{\partial z^2_i} z^3_i  \right] \nonumber \\
     & & +  \text{E}\left[ \left( \frac{1}{2}\frac{\partial^2\mathcal{O}}{\partial z^2_i} \right)^2 z^4_i \right] - \text{E}\left[ A_i \right]^2
\end{eqnarray}
\end{footnotesize}
\noindent where $\text{V}[X]$ and $\text{E}[X]$ corresponds to the standard deviation and mathematical expectation.
Note that $\text{E}[z^2_i]=1$, $\text{E}[z^3_i]=0$ and $\text{E}[z^4_i]=3$ for the gaussian distribution $z_i$. Therefore,
we have

\begin{footnotesize}
\begin{eqnarray}
 \text{V}[A_i] = \left( \frac{\partial \mathcal{O}}{\partial z_i} \right)^2 + \frac{1}{2}\left( \frac{\partial^2\mathcal{O}}{\partial z^2_i}\right)^2.
\end{eqnarray}
\end{footnotesize}

\noindent Similarly, the standard deviation of $B_{ij}$ is:

\begin{footnotesize}
\begin{eqnarray}
 \text{V}[B_{ij}] &=& \text{E}\left[ B^2_{ij} \right] - \text{E}[B_{ij}]^2 = \left(\frac{\partial^2 \mathcal{O}}{\partial z_iz_j}\right)^2
\end{eqnarray}
\end{footnotesize}

Given the fact that the $A_i$ and $B_{ij}$ terms are uncorrelated since they contain independent
gaussian distributions, the standard deviation of $\mathcal{O}$ is simply the sum of
$\text{V}[A_i]$ and $\text{V}[B_{ij}]$, which are positive values. Consequently, the inclusion of second-order derivative terms always increases the size of the PDF-induced uncertainty in the physical observable. Furthermore, including only the second-order diagonal terms generally underestimates the uncertainty. 

\end{appendix}

\end{document}